\documentclass{article}
\usepackage{ijcai22}

\usepackage{times}
\usepackage{soul}
\usepackage{url}
\usepackage[hidelinks]{hyperref}
\usepackage[utf8]{inputenc}
\usepackage[small]{caption}
\usepackage{graphicx}
\usepackage{amsmath}
\usepackage{amsthm}
\usepackage{booktabs}
\usepackage{algorithm}
\usepackage{algorithmic}
\newtheorem{example}{Example}
\newtheorem{theorem}{Theorem}

\usepackage{xspace}
\usepackage{mathtools}
\usepackage[capitalise]{cleveref} 
\usepackage[hidelinks]{hyperref}
\RequirePackage{comment}
\usepackage{xifthen}
\usepackage{arydshln}
\usepackage{xcolor}
\usepackage{amssymb}
\usepackage{dsfont}
\usepackage{bbm}
\usepackage{amssymb}
\usepackage{mathtools}
\usepackage{colortbl}
\usepackage{graphics}
\usepackage{float}
\usepackage{color}
\usepackage{xcolor}
\usepackage{array}
\usepackage{multirow}
\usepackage{enumitem}
\usepackage{tikz}
\usepackage{xifthen}
\makeatletter
\newcommand{\setword}[2]{%
  \phantomsection
  #1\def\@currentlabel{\unexpanded{#1}}\label{#2}%
}
\makeatother
\usepackage{atbegshi,etoolbox}
\makeatletter
\newcommand{\discardpages}[1]{
  \xdef\discard@pages{#1}
  \AtBeginShipout{
    \renewcommand*{\do}[1]{
      \ifnum\value{page}=##1\relax%
        \AtBeginShipoutDiscard
        \gdef\do####1{}
      \fi%
    }%
    \expandafter\docsvlist\expandafter{\discard@pages}
  }%
}

\newcommand{\GG}{\ensuremath{\mathcal G}\xspace}

\newtheorem{proposition}{\bf Proposition}

\newtheorem{lemma}{\bf Lemma}

\newtheorem{definition}{\bf Definition}

\crefname{theorem}{theorem}{\bf Theorem}
\crefname{example}{example}{\bf Example}
\crefname{observation}{observation}{\bf Observation}
\crefname{lemma}{lemma}{\bf Lemma}
\crefname{corollary}{corollary}{\bf Corollary}
\crefname{proposition}{proposition}{\bf Proposition}
\crefname{definition}{definition}{\bf Definition}
\crefname{claim}{claim}{\bf Claim}
\crefname{reductionrule}{reduction rule}{\bf Reduction rule}

\DeclareMathOperator*{\argmaxaa}{arg\,max}
\DeclareMathOperator*{\argminaa}{arg\,min}
\newcommand{\argmi}{\argminaa\limits_}
\newcommand{\argma}{\argmaxaa\limits_}
\newcommand{\instance}{\ensuremath{\mathcal{I}}\xspace}
\newcommand{\suml}{\sum\limits_}
\newcommand{\maxl}{\max\limits_}
\newcommand{\minl}{\min\limits_}
\newcommand{\lb}{\left(}
\newcommand{\rb}{\right)}

\newcommand{\curly}[1]{\ensuremath{\{#1\}}\xspace}
\newcommand{\nn}{\nonumber}
\newcommand{\NPH}{\ensuremath{\mathsf{NP}}-hard\xspace}

\newcommand{\subsum}{{\sc Subset Sum}\xspace}

\newcommand{\setcov}{{\sc Set Cover}\xspace}

\newcommand{\yes}{{\sc Yes}\xspace}

\newcommand{\bud}{\ensuremath{b}\xspace}
\newcommand{\proj}{\ensuremath{P}\xspace}
\newcommand{\voters}{\ensuremath{N}\xspace}
\newcommand{\feasible}{\ensuremath{\mathcal{E}}\xspace}
\newcommand{\prof}{\ensuremath{\mathcal{A}}\xspace}
\newcommand{\ii}{\ensuremath{i}\xspace}

\newcommand{\distinct}{\ensuremath{\hat{n}}\xspace}
\newcommand{\scalel}{\ensuremath{\delta}\xspace}

\newcommand{\hcbp}{\ensuremath{\mathsf{HCBP}}\xspace}
\newcommand{\mmpb}{\ensuremath{\mathsf{MPB}}\xspace}

\newcommand{\lpalgo}{{\sc Ordered-Relax}\xspace}

\newcommand{\opt}{\emph{OPT}}

\newcommand{\lo}{\ensuremath{l_o}\xspace}
\newcommand{\ho}{\ensuremath{h_o}\xspace}
\newcommand{\la}{\ensuremath{l_{\prof}}\xspace}
\newcommand{\ha}{\ensuremath{h_{\prof}}\xspace}
\newcommand{\aof}[1]{%
  \ifthenelse{\isempty{#1}}%
    {\ensuremath{A_i}\xspace}
    {\ensuremath{A_{#1}}\xspace}
}

\newcommand{\uof}[2]{%
  \ifthenelse{\isempty{#1}}%
    {\ensuremath{\operatorname{u_i\!}\left(#2\right)}\xspace}
    {\ensuremath{\operatorname{u_{#1}\!}\left(#2\right)}\xspace}
}
\newcommand{\disuof}[2]{%
  \ifthenelse{\isempty{#1}}%
    {\ensuremath{\operatorname{d_i\!}\left(#2\right)}\xspace}
    {\ensuremath{\operatorname{d_{#1}\!}\left(#2\right)}\xspace}
}
\newcommand{\fdisuof}[2]{%
  \ifthenelse{\isempty{#1}}%
    {\ensuremath{\bud-{\uof{}{#2}}}\xspace}
    {\ensuremath{\bud-{\uof{#1}{#2}}}\xspace}
}
\newcommand{\cof}[1]{\ensuremath{\operatorname{c\!}\left(#1\right)}\xspace}
\newcommand{\cxof}[1]{\ensuremath{\operatorname{c\!}\left(#1\right)x_{#1}}\xspace}

\newcommand{\cxsof}[1]{\ensuremath{\operatorname{c\!}\left(#1\right)x_{#1}^*}\xspace}
\newcommand{\br}{\ensuremath{R}\xspace}
\newcommand{\mmpbr}{\ensuremath{M}\xspace}
\newcommand{\winnerfunction}{\ensuremath{\mathcal{W}}\xspace}
\newcommand{\ruleof}[2]{%
  \ifthenelse{\isempty{#2}}%
    {\ensuremath{\operatorname{#1\!}\left(\instance\right)}\xspace}
    {\ensuremath{\operatorname{#1\!}\left(\instance#2\right)}\xspace}
}
\newcommand{\winners}[2]{%
  \ifthenelse{\isempty{#2}}%
    {\ensuremath{\operatorname{\winnerfunction_{#1}\!}\left(\instance\right)}\xspace}
    {\ensuremath{\operatorname{\winnerfunction_{#1}\!}\left(\instance#2\right)}\xspace}
}
\newcommand{\fullinstance}{\ensuremath{\langle \voters,\proj,c,\bud,\prof \rangle}\xspace}
\newcommand{\firstfrac}[1]{%
  \ifthenelse{\isempty{#1}}%
    {\ensuremath{\frac{|S|-|Y_i|}{|\aof{}|-|Y_i|}}\xspace}
    {\ensuremath{\frac{|S|-|Y_{#1}|}{|\aof{#1}|-|Y_{#1}|}}\xspace}
}
\newcommand{\revfrac}[1]{%
  \ifthenelse{\isempty{#1}}%
    {\ensuremath{\frac{|S|-|\aof{}|}{|S|-|Y_i|}}\xspace}
    {\ensuremath{\frac{|S|-|\aof{#1}|}{|S|-|Y_{#1}|}}\xspace}
}
\newcommand{\frack}[1]{%
  \ifthenelse{\isempty{#1}}%
    {\ensuremath{\frac{|\aof{}|-|Y_i|}{|S|-|Y_i|}}\xspace}
    {\ensuremath{\frac{|\aof{#1}|-|Y_{#1}|}{|S|-|Y_{#1}|}}\xspace}
}

\newcommand{\githublink}{\url{https://github.com/Participatory-Budgeting/maxmin_2022}}

\title{Maxmin Participatory Budgeting}

\author{
Gogulapati Sreedurga
\and
Mayank Ratan Bhardwaj
\And
Y. Narahari
\affiliations
Indian Institute of Science
\emails
\{gogulapatis,mayankb,narahari\}@iisc.ac.in
}

\begin{document}

\maketitle

\begin{abstract}
    Participatory Budgeting (PB) is a popular voting method by which a limited budget is divided among a set of projects, based on the preferences of voters over the projects. PB is broadly categorised as divisible PB (if the projects are fractionally implementable) and indivisible PB (if the projects are atomic). Egalitarianism, an important objective in PB, has not received much attention in the context of indivisible PB. This paper addresses this gap through a detailed study of a natural egalitarian rule, Maxmin Participatory Budgeting (\mmpb), in the context of indivisible PB.  Our study is in two parts: (1) computational (2) axiomatic.  In the first part, we prove that \mmpb is computationally hard and give pseudo-polynomial time and polynomial-time algorithms when parameterized by certain well-motivated parameters. We propose an algorithm that achieves for \mmpb, additive approximation guarantees for restricted spaces of instances and empirically show that our algorithm in fact gives exact optimal solutions on real-world PB datasets. We also establish an upper bound on the approximation ratio achievable for \mmpb by the family of exhaustive strategy-proof PB algorithms. In the second part, we undertake an axiomatic study of the \mmpb rule by generalizing known axioms in the literature. Our study leads to the proposal of a new axiom, \emph{maximal coverage}, which captures fairness aspects. We prove that \mmpb satisfies maximal coverage.
\end{abstract}

\section{Introduction}

Participatory budgeting (PB) is an appealing voting paradigm used for distributing a limited budget (divisible resource such as money, time, etc.) among proposed alternatives (also called projects). It has been deployed in many applications \cite{rocke2014framing} due to its practical relevance. The PB setting in which projects are fractionally implementable is called divisible PB whereas the one in which projects are indivisible, each having a certain cost, is called indivisible PB \cite{aziz2021participatory}. A PB rule aggregates the preferences of voters and proposes a budget division for the projects. Among the various preference elicitation methods, approval votes (each agent specifies a subset of projects to be funded) have received much attention in the PB literature due to their cognitive simplicity \cite{benade2018efficiency}.

In approval-based PB, one way to define the utility of a voter is to base it on the number of approved projects of the voter that are funded \cite{rey2020designing,pierczynski2021proportional}. This notion, however, fails to capture the role played by the size of a project: e.g., suppose the budget is $\$100M$ and a PB rule selects two projects approved by voter $1$,  costing $\$4M$ and $\$6M$, and a project approved by voter $2$ costing $\$90M$. The  utility notion in question gives higher utility to voter $1$ though $90\%$ of the budget is allocated favorably to voter $2$. This immediately motivates the well-studied notion of utility, namely, the total amount allocated to the approved projects of the voter \cite{bogomolnaia2005collective,duddy2015fair,talmon2019framework,aziz2019fair,goel2019knapsack,freeman2021truthful}. We work with this utility notion.

Major goals that are usually sought in voting rules include proportionality, individual excellence, and diversity \cite{faliszewski2017multiwinner}. Proportionality assures that every group of voters receives an allocation proportional to its size and is well studied in the PB literature  \cite{aziz2018proportionally,aziz2021proportionally,pierczynski2021proportional,fairstein2021proportional}. Individual excellence refers to maximizing the sum of utilities of all agents and is hence achieved by the well studied utilitarian rules \cite{talmon2019framework,goel2019knapsack,freeman2021truthful}.
Both proportionality and individual excellence are majority-focused rules. Diversity, on the other hand, is a fairness notion that aims to give representation to every voter, irrespective of whether she belongs to majority or minority. 

Diversity is well achieved by egalitarian rules. For example, suppose a budget of \$50M is to be allocated to school construction projects in three villages, X, Y, and Z, with populations of 10000, 6000, and 2000, respectively. Village X proposes schools at localities $\curly{X_1,X_2,X_3,X_4}$ costing  $\curly{\$10M,\$20M,\$20M,\$60M}$, Y proposes schools at $\curly{Y_1,Y_2,Y_3}$ costing $\curly{\$14M,\$14M,\$16M}$, and Z proposes a school at $Z_1$ costing \$6M respectively. Suppose each voter in a village approves all and only the projects proposed by the village. Utilitarian rules select $\curly{X_1,X_2,X_3}$. Majority focused rules ignore village Z. However, an egalitarian solution $\curly{X_1,X_2,Y_1,Z_1}$ has a more diverse representation and thus promotes universal literacy.

{\bf Contributions and Outline}. Egalitarian rules have been studied in the PB context by Aziz and Stursberg \shortcite{aziz2014generalization}, Aziz et al. \shortcite{aziz2019fair}, Airiau et al. \shortcite{airiau2019portioning}, and Tang et al. \shortcite{tang2020price}. All these works focus only on divisible PB. In the context of indivisible PB, egalitarian rules are studied only for a very restricted special case, multi-winner voting, where the budget is $k$ and the projects cost $1$ each (e.g., Brams et al. \shortcite{brams2007minimax}). Thus, surprisingly, egalitarian rules for indivisible PB have not been studied systematically, barring a case-study by Laruelle \shortcite{laruelle2021voting} that experimentally evaluates a sub-optimal greedy algorithm for an egalitarian objective. In this paper, we address this important gap by systematically studying a  popular egalitarian objective, maxmin, for indivisible PB (\mmpb).

Our choice of maxmin for a detailed study is motivated by the natural appeal, simplicity, and understandability of the maxmin objective in social choice contexts. The objective does bring with it a couple of limitations. The first limitation is applicable not only to maxmin but to all egalitarian rules: presence of outlier voters affects the outcome adversely. However, in many real-life PB elections, it is seen that the voters can be partitioned into a small number (10 to 20) of buckets based on their preferences (e.g., Poland (2018), Aleksandrow (2019) PB elections \cite{stolicki2020pabulib}) and outliers are usually not present. The second limitation is that the utility of a voter with inexpensive approval vote is low by default. However, lower cost of the approval vote is often an indication that the voter is not excited by the project proposals, justifying the low utility. Thus, in our view, these two apparent limitations do not take away the importance and relevance of the maxmin objective. In fact, in real-life, maxmin is found to be preferred over the existing proportional PB rules by the voters \cite{rosenfeld2021should}.\footnote{Considers cardinal utilities. Other preferred rules are either unfair or computationally harder}

Our study of \mmpb for indivisible PB is in two parts: (1) computational (2) axiomatic.  In the first part (\Cref{sec: computation}),
we prove that \mmpb is computationally hard and give pseudo-polynomial time and polynomial-time algorithms when parameterized by certain well-motivated parameters. We propose an algorithm that achieves, for \mmpb, additive approximation guarantees for some restricted spaces and empirically show that our algorithm in fact gives exact optimal solutions on real-world PB datasets. We also establish an upper bound on the approximation ratio achievable for \mmpb by the family of exhaustive strategy-proof PB algorithms. In the second part (\Cref{sec: axioms}), we undertake an axiomatic study of the \mmpb rule by generalizing known axioms in the literature. Our study leads us to propose a new axiom, \emph{maximal coverage}, which captures the fairness notion diversity.  We prove that \mmpb satisfies maximal coverage and achieves diversity.

\section{Notations}\label{sec: prelims}
Let $\voters = \curly{1,\ldots,n}$ be the set of voters and $\proj = \curly{p_1,\ldots,p_m}$ be the set of projects. Let $c: \proj \to \mathbb{N}$ be the cost function and $\bud \in \mathbb{N}$ be the total budget available where $\mathbb{N}$ is the set of natural numbers. The approval vote profile of all voters is represented by $\prof$ where each $\aof{} \in \prof$ is the set of projects approved by the voter \ii. An instance \instance of participatory budgeting is \fullinstance. With a slight abuse of notation, we represent the cost of a set $S$ of projects, $\sum_{p \in S}{\cof{p}}$, by $\cof{S}$. A set $S$ is said to be feasible if $\cof{S} \leq \bud$. A feasible approval vote \aof{} is also called a \emph{knapsack vote}. Let \feasible represent the set of all feasible subsets of projects.

Given an instance \instance, a \textbf{participatory budgeting (PB) rule $\operatorname{\br}$} outputs a set of feasible subsets of projects, i.e., $\ruleof{\br}{} \subseteq \feasible$. A \textbf{PB algorithm} is a rule such that for all \instance, $|\ruleof{\br}{}| = 1$.

The utility of a voter \ii from a set of projects $S$ is defined as the amount of money allocated to the projects approved by \ii, i.e., $\uof{}{S} = \cof{S \cap \aof{}}$.

\begin{definition}[Maxmin Participatory Budgeting (\mmpb)]
Given a PB instance \instance, the $\boldsymbol{\mmpb}$ \textbf{rule} outputs a set of all subsets $S \subseteq \proj$ such that $$S \in \argma{S \in \feasible}{\minl{i \in \voters}{\uof{}{S}}}.$$ 
\end{definition}
For ease of presentation, we call the objective optimized by the \mmpb rule the $\boldsymbol{\mmpb}$ \textbf{objective} or simply $\boldsymbol{\mmpb}$. Given a PB instance \instance and a score $s$, the problem of determining if there exists $S \in \feasible$ such that $\min_{i \in \voters}{\uof{}{S}} \geq s$ is called the \textbf{decision version of} $\boldsymbol{\mmpb}$.

\section{Maxmin PB: Computational Results}\label{sec: computation}
We first prove the hardness of \mmpb and present some tractable special cases based on certain well-motivated parameters. We then give an approximation algorithm for \mmpb and show empirically that, in fact, it gives exact optimal solutions on real-world PB datasets. We conclude by establishing an upper bound on the approximation ratio achieved for \mmpb by any exhaustive strategy-proof PB algorithm. We start by formulating \mmpb as the following \setword{integer linear program}{word: ILP} where each variable corresponds to selection of a project.
\begin{align}
    \nn
    \max\; &q\\
    \label{eq: ipl1}
    \text{subject to }&q \leq \suml{p \in \aof{}}{\cxof{p}} \quad \forall i \in \voters\\\nn
    &\suml{p \in \proj}{\cxof{p}} \leq \bud\\
    \label{eq: ilp2}
    &x_p \in \curly{0,1} \quad \forall p \in \proj\\\nn
    &q \geq 0
\end{align}
\subsection*{NP-Hardness of Maxmin PB}\label{sec: nph}
\begin{theorem}\label{the: strongnph}
The decision version of \mmpb is strongly \NPH.
\end{theorem}
\begin{proof}
We reduce the  \setcov problem to our problem. Given a set $U$ of elements, a collection $S$ of subsets of $U$, and a positive integer $k$, the \setcov problem is to find if there exists $F \subseteq S$ such that $\cup_{P \in F}P = U$ and $|F| = k$. It is known to be strongly \NPH \cite{garey1979computers}. Given an instance $\langle U,S,k \rangle$ of \setcov, we construct an \mmpb instance as follows: For each $C \in S$, create a project $p_c$ with unit cost. For each $a \in U$, create a voter $i$ with $A_i = \curly{p_c: i \in C}$. Set $\bud = k$ and $s = 1$. We claim that both these instances are equivalent.

To prove the correctness of our claim, first assume that we are given a \yes instance of \setcov. That is, there exists $F \subseteq S$ such that $|F|=k$ and all elements of $U$ are covered in $F$. The corresponding set of projects $\curly{p_c: C \in F}$ is feasible since $|F| = k$. Every voter has at least one approved project selected and hence the minimum utility of a voter is at least $1$. Thus, it is a \yes instance of our \mmpb problem. Likewise, if we assume that the reduced instance is a \yes instance, we have a set of $k$ projects such that at least one approved project of each voter is selected. This implies that the given instance is \yes instance of \setcov and completes the proof. 
\end{proof}
\subsection{Tractable Special Cases}\label{sec: tract}
Here, we discuss some tractable special cases of \mmpb.
\subsubsection{Constant number of projects}\label{sec: smallprojects}
We look at this parameter, since, in many scenarios, there can be an upper bound on the number of projects that can be funded due to logistic reasons. The following proposition follows from the fact that solving the \ref{word: ILP} is in FPT when parameterized by the number of variables \cite{lenstra1983integer}.
\begin{proposition}\label{prop: projects}
\mmpb can be solved in polynomial time when the number of projects is constant.
\end{proposition}
\subsubsection{Constant number of distinct votes}\label{sec: smalln}
In many real-world scenarios, though the number of voters is large, the number of distinct votes is small. That is, the set of voters can be partitioned into a small number of equivalence classes such that all voters in an equivalence class approve the same set of projects. For example, the 2018 PB elections held in Powazki (Warsaw, Poland) had 3482 voters, out of which there were only 16 distinct approval votes. In fact, it has been found that in many real-life PB datasets \cite{stolicki2020pabulib}, the number of distinct votes is less than $20\%$ of the number of voters. Information on some such datasets is included at \githublink.

\begin{theorem}\label{the: weaknph}
The decision version of \mmpb is weakly \NPH when the number of distinct votes is constant.
\end{theorem}
\begin{proof}
We reduce the \subsum problem to our problem. Given an integer $Z$ and a set of integers $X= \{x_1,x_2,\ldots,x_n\}$, the \subsum problem is to determine if there exists a subset $X' \subseteq X$ such that $\sum_{x \in X'}{x} = Z$. It is known to be weakly \NPH \cite{garey1979computers}. Given an instance $\langle Z,X \rangle$ of \subsum, we construct an \mmpb instance as follows: For each $x_i \in X$, create a project $p_i$ with cost $x_i$. Create a single voter who approves all the projects. Set $\bud = s = Z$. We claim that both these instances are equivalent.

To prove the correctness, first assume that we are given a \yes instance of \subsum. That is, there exists $X' \subseteq X$ such that  $\sum_{x \in X'}{x} = Z$. The corresponding set of projects $\curly{p_i: x_i \in X'}$ is feasible since $\bud = Z$ and the utility of agent is the total cost of this set and is exactly $Z$. Likewise, if we assume that the reduced instance is a \yes instance, we have a set of projects whose total cost is $Z$. This implies that the given instance is \yes instance of \subsum and completes the proof.
\end{proof}
From \Cref{the: strongnph}, it is known that it is impossible to have a pseudo-polynomial time algorithm if the number of distinct votes is large. We provide a  pseudo-polynomial time algorithm to solve \mmpb when the number of distinct votes is small.

\begin{theorem}
\mmpb can be solved in pseudo-polynomial time when the number of distinct votes is constant.
\end{theorem}
\begin{proof}
Let the number of distinct votes be \distinct. Let $A_1,\ldots,A_{\distinct}$ represent these distinct votes. We propose a dynamic programming algorithm. Construct a $\distinct\!+\!2$ dimensional binary matrix $Q$ such that $Q(i,j,u_1,\ldots,u_{\distinct})$ takes a value $1$ if and only if there exists a subset $S \subseteq \curly{p_1,\ldots,p_i}$ such that $\cof{S} \leq j$ and $\cof{S \cap A_t} = u_t$ for all $t \in [\distinct]$. Here, $i$ takes values from $1$ to $m$, $j$ takes values from $0$ to \bud, and the remaining entries take values from $0$ to $j$. Let the collection of all such $\distinct\!+\!2$ sized tuples be $X$. We fill the first row of the matrix as follows:
\vspace*{-\baselineskip}
\begin{align*}
       Q(1,j,u_1,\ldots,u_{\distinct})  \!=\! \begin{cases} 
      1 & \text{if }j \geq \cof{p_1}, u_t \in \curly{0,\cof{p_1}}, \\&u_t = 0 \!\!\iff\!\! p_1 \notin A_t \;\forall t \in [\distinct] \\
      0 & \text{otherwise} 
   \end{cases}
\end{align*}
Now, we fill the matrix recursively as follows:
\begin{align*}
    Q(i,j,u_1,\ldots,u_{\distinct}) = \max\{&Q(i-1,j,u_1,\ldots,u_{\distinct}),\\&Q(i\!-\!1,j\!-\!\cof{p_i},v_1,\ldots,v_{\distinct})\}
\end{align*}
where for all $t\! \in\! [\distinct]$, $v_t$ is $u_t$ if $p_i \notin A_t$ and $u_t\! -\! \cof{p_i}$ otherwise. We know that there are $|X|$ entries in our matrix. The solution of \mmpb is as follows:
\begin{align*}
    \max\limits_{(m,\bud,u_1,\ldots,u_{\distinct}) \in X}{\lb Q(m,\bud,u_1,\ldots,u_{\distinct})\cdot\minl{t \in [\distinct]}{u_t}\rb}
\end{align*}
There are at most $m(\bud+1)^{\distinct+1}$ tuples in $X$ and computing each entry of our matrix takes constant time. The computation of \mmpb solution from the matrix takes $O(\distinct \bud^{\distinct}\log\bud)$ time. The total running time is $O(m\distinct \bud^{\distinct}\log\bud)$, which is pseudo-polynomial if \distinct is constant. Correctness follows from the definition of $Q$.
\end{proof}
\vspace*{-\baselineskip}
\subsubsection{Constant scalable limit}\label{sec: gcd}
We introduce a new natural parameter, \textit{scalable limit}, that is reasonably small in several real-world PB elections.
\begin{definition}[Scalable Limit]
Given a PB instance \fullinstance, we refer to the ratio $\frac{\max(\cof{p_1},\ldots,\cof{p_m})}{\tiny{GCD}(\cof{p_1},\ldots,\cof{p_m},\bud)}$ as \textbf{scalable limit}, denoted by \scalel.
\end{definition}
Often in many real-world settings, the costs of the projects and the budget are expressed as multiples of some large value. For example, suppose a budget of $10$ billion dollars is to be distributed among a set of projects costing hundreds of millions each. That is, the cost of each project is a multiple of $\$100$M. This PB instance could be scaled down by dividing the costs and budget with $\$100$M to derive a new instance with a budget of $100$. If the cost of the most expensive project originally was $\$900$M, it would now cost $9$ in the scaled down instance. This number $9$ is what we call the scalable limit. In other words, the scalable limit of an instance is the cost of the most expensive project after scaling down the costs and budget to values as low as possible. This parameter takes quite low values in many real-world PB election datasets, e.g., Boston, New York District 8, Seattle District 1 (2019) etc. (see \url{https://pbstanford.org/}).

From \Cref{the: weaknph}, we know that \mmpb is not poly-time solvable even if the number of distinct votes is small. We now prove that, if the scalable limit is also small in conjunction with  the number of distinct votes being small, then \mmpb is poly-time solvable.

\begin{theorem}\label{the: scalablelimit}
\mmpb can be solved in polynomial time when the number of distinct votes and the scalable limit are constant.
\end{theorem}
\begin{proof}
Let $\distinct$ be the number of distinct votes and let $A_1,\ldots,A_{\distinct}$ represent these distinct votes. Divide the costs and budget of the instance by $\tiny{GCD}(\cof{p_1},\ldots,\cof{p_m},\bud)$ to obtain a new instance $\instance'$ with costs $c'$ and budget $\bud'$. Clearly, $\instance'$ has the same optimal \mmpb solution as that of \instance.

It is known that the problem of solving an ILP is in FPT when parameterized by the number of constraints and the highest value in coefficient matrix \cite{ganian2019solving}. We modify the \ref{word: ILP} for \mmpb by replacing \voters with $[\distinct]$ and $c$ with $c'$. Now, the highest value in the coefficient matrix is $\max(c'(p))$, i.e., \scalel, and the number of constraints is \distinct+1. Clearly, the modified ILP is equivalent to the initial ILP and the theorem follows.
\end{proof}
\vspace*{-\baselineskip}
\subsection{A High Performing PB Algorithm}\label{sec: algo}
In this section, we first introduce a family of PB algorithms called as Ordered-Fill algorithms. We then propose an algorithm, called \lpalgo, from this family, that is based on LP-rounding. We prove that it achieves approximation guarantees for \mmpb for some restricted spaces of instances. We show empirically that it provides exact optimal solutions for \mmpb on real-world PB datasets.

\begin{definition}[\textbf{\textit{Ordered-Fill Algorithms}}] Given a participatory budgeting instance \instance, an ordered-fill algorithm w.r.t. a complete order $\succ$ over \proj selects the projects in the decreasing order of their ranks in $\succ$ until the next ranked project does not fit within the budget.\footnote{Greedy rules \cite{talmon2019framework} are ordered-fill algorithms where $\succ$ is based on utility from each \textit{affordable} project.}
\end{definition}
\begin{example}\label{eg: orderedfillalgo}
Consider an instance where $\proj = \curly{p_1,p_2,p_3}$, $\bud = 4$, $\cof{p_1} = \cof{p_3} = 2$, and $\cof{p_2} = 3$. An ordered-fill algorithm w.r.t. $p_1 \succ p_2 \succ p_3$ outputs $\curly{p_1}$.\footnote{Note that the outcome is not maximal since $\curly{p_1,p_3} \in \feasible$.}
\end{example}

We consider the LP-relaxation of \ref{word: ILP} for \mmpb objective by relaxing \cref{eq: ilp2} to $0 \leq x_p \leq 1$.
Consider the following LP-rounding algorithm: Solve the relaxed LP to get $(q^*,x^*)$. Let $S = \phi$ be the initial outcome. Add the project with the highest value of $\cxsof{p}$ to $S$, followed by the one with the second highest value, and so on, till the next project does not fit. Call this algorithm \lpalgo.
\begin{lemma}\label{lem: algobound}
For any PB instance $\instance$, \lpalgo outputs a set $S \in \feasible$ such that $ALG \geq \opt - \frac{|A_j \setminus S|}{|S \setminus A_j|}\cdot(\bud - \opt)$ where $j = \argmi{i \in \voters}{\cof{\aof{} \cap S}}$, $ALG = \cof{\aof{j} \cap S}$, and $\opt$ is the minimum utility in the optimal solution for \mmpb.
\end{lemma}
\vspace*{-\baselineskip}
\begin{proof}
Let $S$ be the outcome of \lpalgo, $j = \argmi{i \in \voters}{\cof{\aof{} \cap S}}$, and $\eta = \frac{|A_j \setminus S|}{|S \setminus A_j|}$.

Let the solution of the relaxed LP be $(q^*,x^*)$. Let $Y_i = S \cap \aof{i}$ for each voter $i$. Since $S \setminus Y_i \subseteq \proj \setminus \aof{i}$, for each $i$,
\begin{align}
    \nn
    \suml{p \in S \setminus Y_i}{\cxsof{p}} &\leq \suml{p \in \proj \setminus \aof{}}{\cxsof{p}}\\
    \label{eq: algo1}
    \suml{p \in S \setminus Y_i}{\cxsof{p}} &\leq \bud - \suml{p \in \aof{}}{\cxsof{p}}
\end{align}
From the design of this algorithm, every project $p$ not in $S$ has $\cxsof{p}$ not more than that of projects in $S$. \begin{align}
    \nn \forall p \in S \setminus Y_i \quad \cxsof{p} &\geq \maxl{p' \in \aof{} \setminus Y_i}{\cxsof{p'}}\\\nn
    &\geq \frac{\suml{p' \in \aof{} \setminus Y_i}{\cxsof{p'}}}{|\aof{}| - |Y_i|}\\\nn
    \suml{p \in S \setminus Y_i}{\cxsof{p}} &\geq \firstfrac{} \suml{p \in \aof{} \setminus Y_i}{\cxsof{p}}
\end{align}
Since $x_p^* \leq 1 \; \forall p$ and $\uof{}{S} = \suml{p \in Y_i}{\cof{p}}$, $\suml{p \in Y_i}{\cxsof{p}} \leq \uof{}{S}$.
\vspace*{-\baselineskip}
\begin{align}
    \label{eq: algo2}
    \suml{ p \in S \setminus Y_i}{\cxsof{p}} \geq \firstfrac{} \lb \suml{p \in \aof{}}{\cxsof{p}} -  \uof{}{S} \rb
\end{align}
From \cref{eq: algo1} and \cref{eq: algo2}:
\begin{align}
    \nn
    \firstfrac{} \lb \suml{p \in \aof{}}{\cxsof{p}} -  \uof{}{S} \rb \leq \bud - \suml{p \in \aof{}}{\cxsof{p}}
\end{align}
\vspace*{-\baselineskip}
\begin{align}
    \nn
    \suml{p \in \aof{}}{\cxsof{p}} &\leq \frac{\bud+\firstfrac{}\;\uof{}{S}}{\firstfrac{}+1}\\
    \label{eq: algo3}
    \suml{p \in \aof{}}{\cxsof{p}} &\leq \frac{\eta\bud+ALG}{1+\eta}
\end{align}
Since the optimal solution also belongs to the feasible region of the relaxed LP, we know that $\opt \leq q^*$. From \cref{eq: ipl1}, we have $q^* \leq \sum_{p \in \aof{j}}{\cxsof{p}}$. Combining these observations with \cref{eq: algo3}, we get,
\begin{align}
    \nn
    \opt &\leq \frac{\eta\bud+ALG}{1+\eta}\\\nn
    ALG &\geq \opt - \eta(\bud - \opt)
\end{align}
This proves the result.
\end{proof}

Given an instance \instance, let \lo and \ho denote respectively the minimum and maximum cardinalities of the outputs produced by all ordered-fill algorithms on $\instance$.
\begin{lemma}\label{lem: loho}
For any instance \instance, \lo and \ho are polynomial-time computable.
\end{lemma}
\begin{proof}
Let $D$ and $A$ respectively denote the outputs of ordered-fill algorithms when the projects are arranged in the non-increasing order and the non-decreasing order of their costs. 
Let $O$ denote the output of any other ordered-fill algorithm. We claim that $|D| \leq |O| \leq |A|$.

\textbf{Part 1 :} $|D| \leq |O|$\\
For the sake of contradiction, let us assume that $|O| < |D|$.
\begin{figure}[ht]
  \centering
  \includegraphics[width=0.63\linewidth]{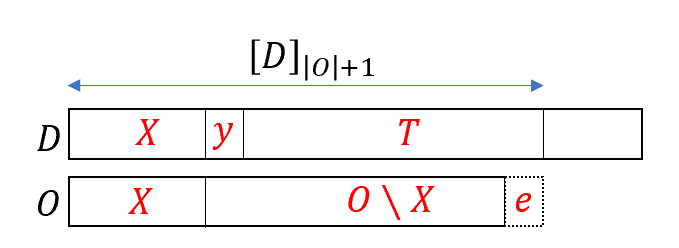}
  \caption{Illustration for $|D| \leq |O|$}
  \label{fig:lleqo}
\end{figure}

Let $X = D \cap O$ be the set of projects selected in both $O$ and $D$. If $D \setminus X = \phi$, our assumption is wrong and the first part of the proof is complete. 

Consider the case where $D \setminus X \neq \phi$. Let $y$ be the project with the highest cost in $\proj \setminus X$. That is,
\begin{align}
    \label{wysx}
    \cof{y} \geq \cof{p} \quad \forall p \in \proj \setminus X
\end{align}
By the definition of $D$,
\begin{align}
    \nn
    y \in D \setminus X
\end{align}
Since $|O| < |D|$, there are at least $|O|+1$ projects in $D$. Hence, there are at least $|O|-|X|$ projects in $D \setminus (X \cup \{y\})$. Let the set of costliest  $|O|-|X|$ projects in $D \setminus (X \cup \{y\})$ be denoted by $T$. By the definition of $D$, $T$ is the set of the costliest $|O| - |X|$ projects in $\proj \setminus (X \cup \{y\})$. Since $y \in D$ and $y \notin X$, $y \notin O$. Thus, $O \setminus X$ is also a set of exactly $|O| - |X|$ projects in $\proj \setminus (X \cup \{y\})$. Therefore,
\begin{align}
    \label{tox}
    \cof{T} \geq \cof{O \setminus X}
\end{align}
As $\{X \cup \{y\} \cup T\} \subseteq D$,
\begin{align}
    \label{xytb}
    \cof{X} + \cof{y} + \cof{T} \leq \bud
\end{align}
Let $e$ be the project where the algorithm stopped adding projects to $O$.
\begin{align}
    \nn
    \cof{O} + \cof{e} &> \bud\\
    \label{xoxeb}
    \cof{X} + \cof{O \setminus X} + \cof{e} &> \bud
\end{align}
Since $e \in \proj \setminus O$, $e \in \proj \setminus X$, from (\ref{wysx}),
\begin{align}
    \label{wewy}
    \cof{y} \geq \cof{e}
\end{align}
From \cref{xytb} and \cref{xoxeb},
\begin{align}
    \nn
    \cof{X} + \cof{y} + \cof{T} &< \cof{X} + \cof{O \setminus X} + \cof{e}\\\nn
    \cof{T} + \cof{y} &< \cof{O \setminus X} + \cof{e}
\end{align}
From \cref{wewy},
\begin{align}
    \nn
    \cof{O \setminus X} > \cof{T}
\end{align}
But this contradicts \cref{tox}, hence contradicting our assumption that $|O| < |D|$. Thus, $|D| \leq |O|$.\\
~\\
\textbf{Part 2 :} $|O| \leq |A|$\\
For the sake of contradiction, let us assume that $|O| > |A|$.
\begin{figure}[ht]
  \centering
  \includegraphics[width=0.63\linewidth]{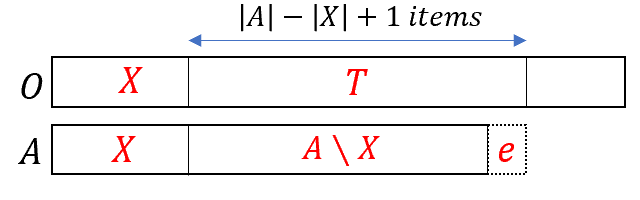}
  \caption{Illustration for $|O| \leq |A|$}
  \label{fig:oleqh}
\end{figure}

Let $X = A \cap O$ be the set of projects selected in both $O$ and $A$. Let $e$ be the first project where the algorithm stopped adding projects to $A$.
\begin{align}
    \nn
    \cof{A} + \cof{e} &> \bud\\
    \label{xaxeb}
    \cof{X} + \cof{A \setminus X} + \cof{e} &> \bud
\end{align}
Since $|O| > |A|$, $|O \setminus X| > |A|-|X|$. Consider a set, $T$, of any $|A|-|X|+1$ projects in $O \setminus X$. From the definition of $A$, $(A \setminus X) \cup \{e\}$ is the set of the least expensive $|A|-|X|+1$ projects in $\proj \setminus X$.
\begin{align}
    \label{tax}
    \cof{A \setminus X} + \cof{e} \leq \cof{T}
\end{align}
Since $T \cup X \subseteq O$,
\begin{align}
    \label{xtb}
    \cof{X} + \cof{T} \leq \bud
\end{align}
From \cref{xaxeb} and \cref{xtb},
\begin{align}
    \nn
    \cof{T} &< \cof{A \setminus X} + \cof{e}
\end{align}
This contradicts \cref{tax}, hence contradicting our assumption that $|O| > |A|$. Thus, $|O| \leq |A|$.

Hence, $|D| \leq |O| \leq |A|$. By definition, $\lo = |D|$ and $\ho = |A|$. $D$ and $A$ can be computed in polynomial time by sorting the projects according to their costs.
\end{proof}
Let \la and \ha respectively denote the lowest and highest cardinalities of the approval votes, i.e., $\la = \min_{i \in \voters}{|\aof{}|}$ and $\ha = \max_{i \in \voters}{|\aof{}|}$.

\Cref{lem: algobound} can be used to establish theoretical approximation guarantees for some restricted spaces of instances, e.g., let us look at a family of instances that satisfy a property that we call \textit{High Cardinality Budget Property} (\hcbp) which intuitively is a cardinal extension of all votes being knapsack votes. That is, it requires that the budget is high enough to fund more projects than the number of projects in any single approval vote. In other words, it requires that $\lo > \ha$. Note that this property is natural in scenarios where budget is high enough to accommodate all projects approved by one voter, e.g., when the budget with the federal government is high enough to fund all proposed projects of one county or when each member in the audience approves less number of talks than the total capacity of the workshop, etc. 
For the instances satisfying \hcbp, our algorithm ensures that $ALG \geq \opt - \ha(\bud - \opt)$\footnote{Further, if we define the disutility of a voter $i$ from a set $S$ to be $\bud - \uof{}{S}$ and minimize the maximum disutility, our algorithm achieves $\lb 2 -\frac{1}{\ho} \rb$ approximation for \hcbp instances. Refer Appendix \ref{app: minimax}.}.

As another example, we note that the algorithm yields an optimal solution for the instances which guarantee $\aof{j} \subseteq S$. One such family of instances will be the one that includes instances  satisfying the following condition for all $S \in \feasible$: if there exists some $p \in \proj$ such that $\cof{S}+\cof{p}>\bud$, then all approved projects of the worse-off voter from $S$ are included in $S$. Note that this is a sufficient but not a necessary condition for the algorithm to yield an  optimal solution. We acknowledge that, for some instances, the approximation ratio given by the lemma could be trivial theoretically, but our empirical analysis compensates for this.

\subsubsection{Empirical Analysis}
Though the theoretical guarantee of our algorithm is limited in terms of approximation ratios and the spaces of instances it covers, empirically it is found to exhibit remarkable performance to give {\em exact optimal solutions\/} on all the PB datasets available \cite{stolicki2020pabulib}. The datasets and corresponding results are included at \githublink. This indicates that, although theoretically the worst cases do not allow the approximation ratio to get better, such worst cases are seldom encountered in real-life, thus explaining the excellent performance of our algorithm.

\subsection{Bound on the Performance of Exhaustive and Strategy-Proof PB Algorithms}\label{sec: sp}
It would be useful to know how the requirement of strategy-proofness will degrade the performance of PB algorithms with respect to \mmpb objective. We establish an upper bound on the approximation ratios of exhaustive strategy-proof PB algorithms by invoking an argument used in the mechanism design literature \cite{procaccia2009approximate,caragiannis2010approximation}. We construct a specific profile whose outcome can be derived using strategy-proofness and establish an upper bound on its approximation ratio.
\begin{definition}[Strategy-proofness]\label{def: sp}
An algorithm \GG is said to be strategy-proof, if and only if, for any instance $\instance$,
\begin{align}
    \nn
    \forall i \quad \forall S \subseteq \proj \quad \cof{\aof{} \cap \GG(\aof{},\prof_{-i})} \geq \cof{\aof{} \cap \GG(S,\prof_{-i})}
\end{align}
where $\prof_{-i}$ denotes the vote profile of all voters except $i$ and $\GG(\prof)$ denotes the outcome of $\GG$ at profile \prof.
\end{definition}
That is, if all voters except $i$ continue to approve the same sets, $i$ should not obtain a better outcome by approving any set other than \aof{}. An algorithm is exhaustive if it is impossible to fund an unselected project with the remaining budget.
\begin{definition}[Exhaustiveness]
An algorithm \GG, is said to be exhaustive if and only if for every instance \instance and every $p \notin \GG(\instance)$, it holds that $\cof{p}+ \cof{\GG(\instance)} > \bud$.
\end{definition}
\begin{theorem}\label{the: exsp}
No exhaustive strategy-proof algorithm can achieve an approximation guarantee better than $\frac{2}{3}$ for \mmpb even for instances with only knapsack votes and unit costs.
\end{theorem}
\begin{proof}
Consider an exhaustive strategy-proof algorithm \GG. Consider an instance $\instance_1$ with a budget \bud, $2\bud$ projects each costing $1$, two voters, and a vote profile $\prof_1 = \curly{\aof{1},\aof{2}}$ such that $\aof{1} \cap \aof{2} = \emptyset$ and $\cof{\aof{1}} = \cof{\aof{2}} = \bud$. Let $\GG(\instance_1) = S$. Without loss of generality, assume that $\cof{\aof{1} \cap S} \leq \cof{\aof{2} \cap S}$. Modify the vote profile of $\instance_1$ to get $\instance_2$ as follows: $\prof_2  = \curly{\aof{1},S}$. For $\instance_2$, \GG should output $S$. Else, voter $2$ would have the incentive to misreport $\aof{j}$ instead of the true vote $S$ and get the preferred outcome $S$. Hence, $\GG(\instance_2) = S$. Therefore, the minimum utility of the algorithm for $\instance_2$, $ALG(\instance_2)$, is $\cof{\aof{1} \cap S}$. Let $\opt(\instance_2)$ be the optimal minimum utility for $\instance_2$. That is,
\begin{align}
    \label{eq: ic1}
    \opt(\instance_2) \geq \min{(\cof{\aof{1} \cap X}, \cof{S \cap X})} \quad \forall X \in \feasible
\end{align}
Since \GG is exhaustive, $\cof{S} = \bud$. Consider a set $X \subseteq \proj$ such that $(\aof{1} \cap S) \subseteq X$, $|X \cap (\aof{1} \setminus S)| = \frac{\bud - \cof{\aof{1} \cap S}}{2}$, and $|X \cap (S \setminus \aof{1})| = \frac{\bud - \cof{\aof{1} \cap S}}{2}$. Clearly, $X \in \feasible$.
\begin{align}
    \nn
    \cof{\aof{1} \cap X} = \cof{S \cap X} &= \cof{\aof{1} \cap S} + \frac{\bud - \cof{\aof{1} \cap S}}{2}\\
    \label{eq: ic2}
    &=  \frac{\bud + \cof{\aof{1} \cap S}}{2}
\end{align}
Since $\cof{\aof{1} \cap S} \leq \cof{\aof{2} \cap S}$ and $\cof{S} = \bud$, $\cof{\aof{1} \cap S} \leq \frac{\bud}{2}$. Substituting $\bud \geq 2\cof{\aof{1} \cap S}$ in \cref{eq: ic2}, we get,
\begin{align}
    \tag{From \cref{eq: ic1}}
    \opt(\instance_2) &\geq \frac{3\cof{\aof{1} \cap S}}{2}\\\nn
    \frac{ALG(\instance_2)}{\opt(\instance_2)} &\leq \frac{2}{3}
\end{align}
Hence, the approximation guarantee of the algorithm is at most $\frac{2}{3}$. Clearly in both the instances $\instance_1$ and $\instance_2$, all the votes are knapsack votes and the theorem follows.
\end{proof}

\section{Maxmin PB: An Axiomatic Analysis}\label{sec: axioms}
An axiomatic study of a voting rule provides valuable insights into the characteristics of the voting rule. We now undertake an axiomatic study of the \mmpb rule by exploring several axiomatic properties available in the literature.
Given an instance \instance and a PB rule \br, we say a project $p$ wins if there exists $S \in \ruleof{\br}{}$ such that $p \in S$. Let \winners{\br}{} represent the set of all projects which win, i.e., $\winners{\br}{} = \curly{p \in \proj: \exists S \in \ruleof{\br}{} p \in S}$. Throughout this section, we represent the \mmpb rule by \mmpbr.

First, we examine axioms in the PB literature for rules that output a single feasible subset of projects \cite{talmon2019framework}. We then extend the axioms to rules (like \mmpb) that output multiple feasible subsets. We start by defining all the axioms. The first axiom requires that if a winning project is replaced by a set of projects whose total cost is the same, then at least one project of this set should continue to win.
\begin{definition}[Splitting Monotonicity]\label{def: splitting}
A PB rule \br is said to satisfy splitting monotonicity iff, for any instance \instance, $\winners{\br}{'} \cap P' \neq \emptyset$ whenever $\instance'$ is obtained from \instance by splitting a project $p \in \winners{\br}{}$ into a set of projects $P'$ with $\cof{P'} = \cof{p}$ and changing every $\aof{}$ having $p$ to $(\aof{} \setminus \curly{p}) \cup P'$.
\end{definition}
The next axiom requires that, if we replace a set of winning projects that are approved by exactly same set of voters by a single project having the same total cost, then the single project wins.
\begin{definition}[Merging Monotonicity]\label{def: merging}
A PB rule \br is said to satisfy merging monotonicity iff, for any instance \instance, $p \in \winners{\br}{'}$ whenever $\instance'$ is obtained from \instance as follows: for any $S \in \ruleof{\br}{}$ and $P' \subseteq S$ such that $\aof{} \cap P' \in \curly{P',\emptyset}$ for every voter \ii, merge the projects in $P'$ into a single project $p$ with cost $\cof{P'}$ and change every $\aof{}$ with $P'$ to $(\aof{} \setminus P') \cup \curly{p}$.
\end{definition}
The next axiom requires that no winning project should be dropped if it becomes less expensive.
\begin{definition}[Discount Monotonicity]\label{def: discount}
A PB rule \br is said to satisfy discount monotonicity iff, for any instance \instance, $p \in \winners{\br}{'}$ whenever $\instance'$ is obtained from \instance by reducing the cost of $p \in \winners{\br}{}$ to $\cof{p}-1$.
\end{definition}
The following axiom requires that no winning project should be dropped if the budget increases.
\begin{definition}[Limit Monotonicity]\label{def: limit}
A PB rule \br is said to satisfy limit monotonicity iff, for any instance \instance such that no project in \proj costs $\bud+1$, $\winners{\br}{} \subseteq \winners{\br}{'}$ whenever $\instance'$ is obtained from \instance by increasing the budget to $\bud+1$.
\end{definition}
The next axiom requires that all sets in \ruleof{\br}{} are maximal \cite{aziz2021participatory}.
\begin{definition}[Strong Exhaustiveness]\label{def: stronge}
A PB rule \br is said to satisfy strong exhaustiveness iff, for any instance \instance, $$\forall S \in \ruleof{\br}{}\;\forall p \in \proj\setminus S \quad \cof{S} + \cof{p} > \bud$$
\end{definition}
Next, we introduce a new axiom, weak exhaustiveness, which requires that any winning non-maximal set can be made maximal without compromising on the win.
\begin{definition}[Weak Exhaustiveness]\label{def: weake}
A PB rule \br is said to satisfy weak exhaustiveness iff, for any instance \instance, $$\forall S\!\! \in\!\! \ruleof{\br}{},\;\forall p \in \proj\!\!\setminus\!\!S,\; \cof{S} \!\!+\!\! \cof{p} \leq \bud \!\!\implies\!\! S\cup\curly{p}\!\! \in\!\! \ruleof{\br}{}$$
\end{definition}
\begin{theorem}\label{the: pbaxioms_satisfy}
The \mmpb rule satisfies: (a) splitting monotonicity (b) merging monotonicity (c) weak exhaustiveness.
\end{theorem}
\begin{proof}~\\
\noindent{\textbf{(a) Splitting Monotonicity:}}\\
Let $p \in \winners{\mmpbr}{}$ be split into a set of projects $P'$ as above to produce $\instance'$. For the sake of contradiction, let us assume that $\winners{\mmpbr}{'} \cap P' = \emptyset$. Since $p \in \winners{\mmpbr}{}$, $\exists S \in \ruleof{\mmpbr}{}$ such that $p \in S$. Consider a set of projects $K = (S\setminus \{p\}) \cup P'$. Since $\cof{\aof{} \cap K}$ and $\cof{\aof{} \cap S}$ are same for every voter \ii, the minimum utility from $K$ is equal to that from $S$. From our assumption, for any set $T \in \ruleof{\mmpbr}{'}$, $T \cap P' = \emptyset$ and the minimum utility from $T$ is strictly greater than that from $K$ and $S$. This contradicts the fact that $S \in \ruleof{\mmpbr}{}$.

\noindent{\textbf{(b) Merging Monotonicity:}}\\
Let $S \in \ruleof{\br}{}$ and $P' \subseteq S$ satisfy the condition specified. That is, each voter either approves entire $P'$ or approves no project in $P'$. Let $P'$ be merged into a single project $p$ and the new instance thus produced be $\instance'$. For the sake of contradiction, let us assume that $p \notin \winners{\br}{'}$. Consider the set $K = (S\setminus P') \cup \{p\}$. Since $\cof{\aof{} \cap K}$ and $\cof{\aof{} \cap S}$ are same for every voter \ii, the minimum utility from $K$ is equal to that from $S$. From our assumption, for any set $T \in \ruleof{\mmpbr}{'}$, $p \notin T$ and the minimum utility from $T$ is strictly greater than that from $K$ and $S$. This contradicts $S \in \ruleof{\mmpbr}{}$.

\noindent{\textbf{(c) Weak Exhaustiveness:}}\\
For the sake of contradiction, let us assume that $\exists S \in \ruleof{\mmpbr}{}$ and $p \in \proj\setminus S$ such that $\cof{S} + \cof{p} \leq \bud$. Consider the feasible set $K = S \cup \{p\}$. Consider any arbitrary voter \ii. If $p \in \aof{}$, $\uof{}{K} = \uof{}{S} + \cof{p}$. Else if $p \notin \aof{}$, $\uof{}{K} = \uof{}{S}$. So, the minimum utility of any voter from $K$ is at least that from $S$. Since $S \in \ruleof{\mmpbr}{}$, $K \in \ruleof{\mmpbr}{}$.
\end{proof}
\begin{theorem}\label{the: pbaxioms_nsatisfy}
The \mmpb rule does not satisfy: (a)  discount monotonicity (b) limit monotonicity (c) strong exhaustiveness.
\end{theorem}
\begin{proof}~\\
\noindent{\textbf{(a) Discount Monotonicity:}}\\
Consider an instance $\instance$ with $\proj = \curly{p_1,p_2,p_3,p_4}$ each costing $4$, a budget $12$, and three voters with $\aof{1} = \curly{p_1,p_2}$, $\aof{2} = \curly{p_3}$, and $\aof{3} = \curly{p_4}$. Since $\ruleof{\mmpbr}{}=\{\{p_1,p_3,p_4\}, \{p_2,p_3,p_4\}\}$, $p_2 \in \winners{\mmpbr}{}$. Now consider the case where $\cof{p_2}$ is reduced to $3$ to get new instance $\instance'$. Then, $\ruleof{\mmpbr}{'} = \{\{p_1,p_3,p_4\}\}$ and hence $p_2 \notin \winners{\mmpbr}{'}$.

\noindent{\textbf{(b) Limit Monotonicity:}}\\
Consider an instance $\instance$ with $\proj = \curly{p_1,p_2,p_3,p_4,p_5,p_6}$ costing $\curly{3,1,3,3,3,6}$ respectively, a budget $12$, and four voters with $\aof{1} = \curly{p_1,p_2}$, $\aof{2} = \curly{p_3,p_4}$, $\aof{3} = \curly{p_5}$, and $\aof{4} = \curly{p_6}$. Clearly, $\{p_1,p_3,p_5\} \in \ruleof{\mmpbr}{}$. Therefore, $p_1 \in \winners{\mmpbr}{}$. If \bud is increased to $13$ to get $\instance'$, $\ruleof{\mmpbr}{'} = \{\{p_2,p_4,p_5,p_6\}, \{p_2,p_3,p_5,p_6\}\}$ and $p_1 \notin \winners{\mmpbr}{'}$.

\noindent{\textbf{(c) Strong Exhaustiveness:}}\\
The above example also illustrates the counter example for strong exhaustiveness. Clearly, $S = \{p_1,p_3,p_5\} \in \ruleof{\mmpbr}{}$ but $\cof{p_2}+\cof{S} < 12$.
\end{proof}
We now examine two important axioms from multi-winner voting literature. The first axiom is an analogue of unanimity, which requires that a project approved by all voters needs to win \cite{faliszewski2017multiwinner}.
\begin{definition}[Narrow-top Criterion]\label{def: narrowtop}
A PB rule \br is said to satisfy narrow-top criterion iff, for any instance \instance, $p \in \winners{\br}{}$ whenever $p \in \aof{}$ for every $i\in \voters$.
\end{definition}
\begin{proposition} \label{prop: narrowtop}
The \mmpb rule  does not satisfy narrow-top criterion.
\end{proposition}
\begin{proof}
Our example uses the fact that the utility from an unanimously approved project could be low and selecting it could make the other projects unaffordable. Consider an instance $\instance$ with $\proj = \curly{p_1,p_2,p_3}$ costing $\curly{1,3,3}$ respectively, a budget $6$, and two voters with $\aof{1}\!\! =\!\! \curly{p_1,p_2}$ and $\aof{2}\!\! =\!\! \curly{p_1,p_3}$. Since $\ruleof{\mmpbr}{}\!\!=\!\!\{\{p_2,p_3\}\}$, we have $p_1\!\! \notin\!\! \winners{\mmpbr}{}$ though $p_1 \!\!\in\!\! \aof{}$ for all \ii.
\end{proof}
The next axiom tries to capture diversity in the voting rules \cite{aziz2018egalitarian,brandt2016handbook}.
\begin{definition}[Clone Independence]\label{def: clone}
A PB rule \br is said to satisfy clone independence iff, for any instance \instance, $\ruleof{\br}{}$ does not change when a group of voters, all having exactly the same approval vote $\aof{v}$, is replaced by a single voter with approval vote $\aof{v}$.
\end{definition}
\begin{proposition}\label{prop: clone}
The \mmpb rule satisfies clone independence.
\end{proposition}
The above proposition follows from the fact that redundant votes do not affect the value of maxmin objective. Clearly, this axiom represents diversity, but it is rather narrow. Faliszewski et al. \shortcite{faliszewski2017multiwinner} identified that the fairness notion of diversity does not have a clear axiomatic representation in the literature. We address this gap by introducing a new axiom, called \emph{Maximal Coverage}, to capture diversity in PB as well as in multi-winner voting settings.

Let us define {\em covered voters\/}  as the voters with at least one approved project funded and a {\em redundant project\/} as a project whose removal from an outcome does not change the set of covered voters. To achieve diversity, we need to cover as many voters as possible. Our new axiom ensures that a redundant project is funded only when no more voters can be covered by doing otherwise. For example, while allocating time for plenary talks, the organizers might want to explore as many novel ideas as possible, without eliminating any key  area. They might allocate a time slot to a proposed plenary talk on a not-so-popular topic (hoping to prop up a new area), by dropping 1 out of 4 proposed talks in a popular area. This {\em covers\/} or reaches out to the audience from the new area and broadens the reach of conference. 

\begin{definition}[Maximal Coverage]\label{def: maximal}
A PB rule \br is said to satisfy maximal coverage iff, for any instance \instance, $S \in \ruleof{\br}{}$, $p \in S$ such that $\{j_1:p \in \aof{j_1}\} \subseteq \{j_2:(S\setminus\{p\})\cap \aof{j_2} \neq \emptyset\}$, and $i \in \voters$,
\begin{equation}
    \label{eq: maximal1}
    \winners{\br}{} \cap \aof{} = \emptyset \implies \cof{a} > \bud -\cof{S\setminus\{p\}} \; \forall a \in \aof{}
\end{equation}
\end{definition}
We consider an example to understand this better. Suppose a budget of \$2.25B is to be allocated to projects in two counties, X and Y, with populations of 10000 and 6000, resp. County X proposes projects $\curly{X_1, X_2,X_3}$ costing $\curly{\$0.5B,\$1B,\$1B}$, and County Y proposes projects $\curly{Y_1, Y_2,Y_3}$ costing $\curly{\$0.7B,\$0.7B,\$0.8B}$ resp. Assume each voter approves all and only the projects of her county. A utilitarian rule selects the set $\{\{X_2, X_3\}\}$. Let $p$ be $X_2$. 
If $i = 2$ and $a = Y_1$, then the first part of \cref{eq: maximal1} satisfies but $\cof{Y_1} \leq \bud - \cof{X_3}$. Hence it does not satisfy maximal coverage. If we apply the \mmpb rule, we get $\winners{\mmpbr}{} = \{X_1, X_2, X_3, Y_1, Y_2\}$ and $\ruleof{\mmpbr}{} = \{\{X_1, X_2, Y_1\}, \{X_1, X_2, Y_2\}, \{X_1, X_3, Y_1\}, \{X_1, X_3, Y_2\}\}$. It is notable that \cref{eq: maximal1} holds.
\begin{proposition}\label{prop: maximal}
The \mmpb rule satisfies maximal coverage.
\end{proposition}
\vspace{-3mm}
\begin{proof}
Let $S,p,i,a$ be as stated in \Cref{def: maximal}. Assume $\exists S$  such that the first part of \cref{eq: maximal1} holds. Since $\winners{\mmpbr}{} \cap \aof{} = \emptyset$, $\uof{}{S} = 0$. Hence the minimum utility from any set in $\ruleof{\mmpbr}{}$ is $0$. But, since $\winners{\mmpbr}{} \cap \aof{} = \emptyset$ and $a \in \aof{}$, $\{a\} \notin \ruleof{\mmpbr}{}$. This is possible only if $\cof{a} > \bud$ and, thus, the second part of \cref{eq: maximal1} holds.
\end{proof}

\section{Summary and Future Directions}
In this paper, we have motivated the \mmpb rule for indivisible PB and undertaken  a computational as well as an axiomatic study of the same. On the computational side, we proved  \mmpb is strongly \NPH. We then proved it is weakly \NPH when the number of distinct votes is small and proposed a pseudo-polynomial time algorithm. We introduced a novel parameter, scalable limit, and proved that if it is also small, \mmpb can be solved in polynomial-time. We proposed an LP-rounding algorithm that gives additive approximation guarantees for restricted spaces of instances and showed empirically that it gives  optimal outcomes on real-world PB datasets. We established an upper bound on approximation ratio achievable by the family of exhaustive and strategy-proof algorithms. On the axiomatic side, we undertook a detailed analysis of \mmpb and introduced a new axiom, maximal coverage, to capture diversity in PB and multi-winner voting.

Giving an axiomatic characterization of \mmpb, studying other objectives (for example leximin), establishing the connections between maximal coverage and other axioms in the literature are a few of several interesting directions to pursue.

\section*{Acknowledgments}
Gogulapati Sreedurga gratefully acknowledges the Prime Minister Research Fellowship funded by Government of India. Mayank Ratan Bhardwaj gratefully acknowledges the MHRD funding by Government of India. The authors thank Dr. Neeldhara Misra for suggesting the number of distinct votes as a parameter, and the anonymous reviewers for some helpful feedback.

\bibliographystyle{named}
{\small
\bibliography{ijcai22}}

\begin{thebibliography}{}

\bibitem[\protect\citeauthoryear{Airiau \bgroup \em et al.\egroup
  }{2019}]{airiau2019portioning}
St{\'e}phane Airiau, Haris Aziz, Ioannis Caragiannis, Justin Kruger,
  J{\'e}r{\^o}me Lang, and Dominik Peters.
\newblock Portioning using ordinal preferences: Fairness and efficiency.
\newblock In {\em IJCAI 2019}, pages 11--17, 2019.

\bibitem[\protect\citeauthoryear{Aziz and Lee}{2021}]{aziz2021proportionally}
Haris Aziz and Barton~E Lee.
\newblock Proportionally representative participatory budgeting with ordinal
  preferences.
\newblock In {\em AAAI 2021}, pages 5110--5118, 2021.

\bibitem[\protect\citeauthoryear{Aziz and Shah}{2021}]{aziz2021participatory}
Haris Aziz and Nisarg Shah.
\newblock Participatory budgeting: Models and approaches.
\newblock In {\em Pathways Between Social Science and Computational Social
  Science}, pages 215--236. Springer, 2021.

\bibitem[\protect\citeauthoryear{Aziz and
  Stursberg}{2014}]{aziz2014generalization}
Haris Aziz and Paul Stursberg.
\newblock A generalization of probabilistic serial to randomized social choice.
\newblock In {\em AAAI 2014}, pages 559--565, 2014.

\bibitem[\protect\citeauthoryear{Aziz \bgroup \em et al.\egroup
  }{2018a}]{aziz2018egalitarian}
Haris Aziz, Piotr Faliszewski, Bernard Grofman, Arkadii Slinko, and Nimrod
  Talmon.
\newblock Egalitarian committee scoring rules.
\newblock In {\em IJCAI 2018}, pages 56--62, 2018.

\bibitem[\protect\citeauthoryear{Aziz \bgroup \em et al.\egroup
  }{2018b}]{aziz2018proportionally}
Haris Aziz, Barton~E Lee, and Nimrod Talmon.
\newblock Proportionally representative participatory budgeting: Axioms and
  algorithms.
\newblock In {\em AAMAS 2018}, pages 23--31, 2018.

\bibitem[\protect\citeauthoryear{Aziz \bgroup \em et al.\egroup
  }{2019}]{aziz2019fair}
Haris Aziz, Anna Bogomolnaia, and Herv{\'e} Moulin.
\newblock Fair mixing: The case of dichotomous preferences.
\newblock In {\em Proceedings of the 2019 ACM EC}, pages 753--781, 2019.

\bibitem[\protect\citeauthoryear{Benade \bgroup \em et al.\egroup
  }{2018}]{benade2018efficiency}
Gerdus Benade, Nevo Itzhak, Nisarg Shah, and Ariel~D Procaccia.
\newblock Efficiency and usability of participatory budgeting methods.
\newblock \url{https://www.cs.toronto.edu/~nisarg/papers/pb_usability.pdf},
  2018.

\bibitem[\protect\citeauthoryear{Bogomolnaia \bgroup \em et al.\egroup
  }{2005}]{bogomolnaia2005collective}
Anna Bogomolnaia, Herv{\'e} Moulin, and Richard Stong.
\newblock Collective choice under dichotomous preferences.
\newblock {\em Journal of Economic Theory}, 122(2):165--184, 2005.

\bibitem[\protect\citeauthoryear{Brams \bgroup \em et al.\egroup
  }{2007}]{brams2007minimax}
Steven~J Brams, D~Marc Kilgour, and M~Remzi Sanver.
\newblock A minimax procedure for electing committees.
\newblock {\em Public Choice}, 132(3-4):401--420, 2007.

\bibitem[\protect\citeauthoryear{Brandt \bgroup \em et al.\egroup
  }{2016}]{brandt2016handbook}
Felix Brandt, Vincent Conitzer, Ulle Endriss, J{\'e}r{\^o}me Lang, and Ariel~D
  Procaccia.
\newblock {\em Handbook of computational social choice}.
\newblock Cambridge University Press, 2016.

\bibitem[\protect\citeauthoryear{Caragiannis \bgroup \em et al.\egroup
  }{2010}]{caragiannis2010approximation}
Ioannis Caragiannis, Dimitris Kalaitzis, and Evangelos Markakis.
\newblock Approximation algorithms and mechanism design for minimax approval
  voting.
\newblock In {\em AAAI 2010}, pages 737--742, 2010.

\bibitem[\protect\citeauthoryear{Duddy}{2015}]{duddy2015fair}
Conal Duddy.
\newblock Fair sharing under dichotomous preferences.
\newblock {\em Mathematical Social Sciences}, 73:1--5, 2015.

\bibitem[\protect\citeauthoryear{Fairstein \bgroup \em et al.\egroup
  }{2021}]{fairstein2021proportional}
Roy Fairstein, Reshef Meir, and Kobi Gal.
\newblock Proportional participatory budgeting with substitute projects.
\newblock {\em arXiv preprint arXiv:2106.05360}, 2021.

\bibitem[\protect\citeauthoryear{Faliszewski \bgroup \em et al.\egroup
  }{2017}]{faliszewski2017multiwinner}
Piotr Faliszewski, Piotr Skowron, Arkadii Slinko, and Nimrod Talmon.
\newblock Multiwinner voting: A new challenge for social choice theory.
\newblock {\em Trends in computational social choice}, 74:27--47, 2017.

\bibitem[\protect\citeauthoryear{Freeman \bgroup \em et al.\egroup
  }{2021}]{freeman2021truthful}
Rupert Freeman, David~M Pennock, Dominik Peters, and Jennifer~Wortman Vaughan.
\newblock Truthful aggregation of budget proposals.
\newblock {\em Journal of Economic Theory}, 193:105234, 2021.

\bibitem[\protect\citeauthoryear{Ganian and Ordyniak}{2019}]{ganian2019solving}
Robert Ganian and Sebastian Ordyniak.
\newblock Solving integer linear programs by exploiting variable-constraint
  interactions: A survey.
\newblock {\em Algorithms}, 12(12):248, 2019.

\bibitem[\protect\citeauthoryear{Garey and Johnson}{1979}]{garey1979computers}
Michael~R Garey and David~S Johnson.
\newblock {\em Computers and intractability}, volume 174.
\newblock San Francisco: freeman, 1979.

\bibitem[\protect\citeauthoryear{Goel \bgroup \em et al.\egroup
  }{2019}]{goel2019knapsack}
Ashish Goel, Anilesh~K Krishnaswamy, Sukolsak Sakshuwong, and Tanja Aitamurto.
\newblock Knapsack voting for participatory budgeting.
\newblock {\em ACM Transactions on Economics and Computation (TEAC)},
  7(2):1--27, 2019.

\bibitem[\protect\citeauthoryear{Laruelle}{2021}]{laruelle2021voting}
Annick Laruelle.
\newblock Voting to select projects in participatory budgeting.
\newblock {\em European Journal of Operational Research}, 288(2):598--604,
  2021.

\bibitem[\protect\citeauthoryear{Lenstra~Jr}{1983}]{lenstra1983integer}
Hendrik~W Lenstra~Jr.
\newblock Integer programming with a fixed number of variables.
\newblock {\em Mathematics of operations research}, 8(4):538--548, 1983.

\bibitem[\protect\citeauthoryear{Pierczy{\'n}ski \bgroup \em et al.\egroup
  }{2021}]{pierczynski2021proportional}
Grzegorz Pierczy{\'n}ski, Piotr Skowron, and Dominik Peters.
\newblock Proportional participatory budgeting with additive utilities.
\newblock {\em Advances in Neural Information Processing Systems}, 34, 2021.

\bibitem[\protect\citeauthoryear{Procaccia and
  Tennenholtz}{2009}]{procaccia2009approximate}
Ariel~D Procaccia and Moshe Tennenholtz.
\newblock Approximate mechanism design without money.
\newblock In {\em Proceedings of the 10th ACM conference on electronic
  commerce}, pages 177--186, 2009.

\bibitem[\protect\citeauthoryear{Rey \bgroup \em et al.\egroup
  }{2020}]{rey2020designing}
Simon Rey, Ulle Endriss, and Ronald de~Haan.
\newblock Designing participatory budgeting mechanisms grounded in judgment
  aggregation.
\newblock In {\em KR 2020}, pages 692--702, 2020.

\bibitem[\protect\citeauthoryear{R{\"o}cke}{2014}]{rocke2014framing}
Anja R{\"o}cke.
\newblock {\em Framing citizen participation: Participatory budgeting in
  France, Germany and the United Kingdom}.
\newblock Springer, 2014.

\bibitem[\protect\citeauthoryear{Rosenfeld and
  Talmon}{2021}]{rosenfeld2021should}
Ariel Rosenfeld and Nimrod Talmon.
\newblock What should we optimize in participatory budgeting?: An experimental
  study.
\newblock {\em arXiv preprint arXiv:2111.07308}, 2021.

\bibitem[\protect\citeauthoryear{Stolicki \bgroup \em et al.\egroup
  }{2020}]{stolicki2020pabulib}
Dariusz Stolicki, Stanis{\l}aw Szufa, and Nimrod Talmon.
\newblock Pabulib: A participatory budgeting library.
\newblock {\em arXiv preprint arXiv:2012.06539}, 2020.

\bibitem[\protect\citeauthoryear{Talmon and
  Faliszewski}{2019}]{talmon2019framework}
Nimrod Talmon and Piotr Faliszewski.
\newblock A framework for approval-based budgeting methods.
\newblock In {\em AAAI 2019}, pages 2181--2188, 2019.

\bibitem[\protect\citeauthoryear{Tang \bgroup \em et al.\egroup
  }{2020}]{tang2020price}
Zhongzheng Tang, Chenhao Wang, and Mengqi Zhang.
\newblock Price of fairness in budget division for egalitarian social welfare.
\newblock In {\em ICCOA 2020}, pages 594--607. Springer, 2020.

\end{thebibliography}

\appendix
\section{Minimizing the maximum disutility}\label{app: minimax}
The disutility notion we consider is $\bud - \uof{}{S}$. This notion effectively handles the scenarios where the voter, being a taxpayer and contributor to the budget, pays a fixed amount irrespective of the cost of its approval vote. The voter expects superior projects (which it believes are worth the tax paid) to be proposed and also to be selected. Unsatisfactory proposals or non-funded approved projects make a voter unhappy. In such a scenario, it is reasonable to assume that every dollar of the budget not used for approved projects causes disutility to the voter since it amounts to a waste.

From the objective function, it is clear that minimizing maximum disutility is equivalent to maximizing minimum utility. However, the approximation results do not transfer. To derive the approximation guarantee of our algorithm \lpalgo, we first formulate the ILP.

\begin{align}
    \nn
    \min\; &q\\
    \label{eq: ipl1mm}
    \text{subject to }&q \geq \bud - \suml{p \in \aof{}}{\cxof{p}} \quad \forall i \in \voters\\\nn
    &\suml{p \in \proj}{\cxof{p}} \leq \bud\\
    \label{eq: ilp2mm}
    &x_p \in \curly{0,1} \quad \forall p \in \proj\\\nn
    &q \geq 0
\end{align}

We consider the LP-relaxation for minimax objective by relaxing \cref{eq: ilp2mm} to $0 \leq x_p \leq 1$. Now, again consider the same \lpalgo algorithm: Solve the relaxed LP to get $(q^*,x^*)$. Let $S = \phi$ be the initial outcome. Add the project with the highest value of $\cxsof{p}$ to $S$, followed by the one with the second highest value and so on till the next project does not fit.

\begin{theorem}
The algorithm \lpalgo achieves an approximation guarantee of $\left(\;2-\frac{1}{\ho}\right)$ for \hcbp instances for the objective of minimizing the maximum disutility.
\end{theorem}
\begin{proof}
We obtain the below step following the steps exactly in the proof of \Cref{lem: algobound}:
\begin{align}
    \nn
    \suml{p \in \aof{}}{\cxsof{p}} &\leq \frac{\bud+\firstfrac{}\;\uof{}{S}}{\firstfrac{}+1}\\
    \label{app: eq1mm}
    \lb \frac{\bud - \uof{}{S}}{1 + \lb \frack{} \rb} \rb &\leq \bud - \suml{p \in \aof{}}{\cxsof{p}}
\end{align}
Let $j = \arg\max_{i}{(\fdisuof{}{S})}$ and the maximum disutility achieved by \lpalgo be $ALG = \fdisuof{j}{S}$. Let $\opt\;$ be the value of the maximum disutility in the optimal solution of \mmpb objective. Since the optimal solution also belongs to the feasible region of the relaxed LP, we know that $q^* \leq \opt$. From \cref{eq: ipl1mm}, we know that $q^* \geq \bud - \suml{p \in \aof{j}}{\cxsof{p}}$. Combining these, we have,
\begin{align}
    \nn
    \opt &\geq \lb \frac{\bud - \uof{}{S}}{1 + \lb \frack{} \rb} \rb\\
    \nn
    ALG &\leq \lb 1 + \lb \frack{} \rb \rb\opt\\
    \nn
    &\leq \lb 2 - \revfrac{j} \rb\opt
\end{align}
Since $Y_j \subseteq S$, $|S| - |Y_j| \leq |S| \leq \ho$. Since the instance satisfies \hcbp, $|S| \geq \lo > \ha \geq |\aof{j}|$. Therefore $|S| > |\aof{j}|$, which implies $|S| - |\aof{j}| \geq 1$. Hence, $\revfrac{j} \geq \frac{1}{\ho}$ and $ALG \leq \lb 2 - \frac{1}{\ho} \rb \cdot \opt$
\end{proof}
\end{document}